\documentclass[%
 reprint,
superscriptaddress,
nofootinbib,
 amsmath,amssymb,
 aps,
epja,
]{revtex4-1}
\newcommand{\kaonangle}{$\cos\theta_\mathrm{CM}^{K}$}
\newcommand{\kaon}{$K^0$}
\newcommand{\pion}{$\pi^0$}
\newcommand{\mysigma}{$\Sigma^0$}
\newcommand{\mygamma}{$\gamma $}
\usepackage{xcolor}
\usepackage{siunitx}
\usepackage{graphicx}% Include figure files
\usepackage{dcolumn}% Align table columns on decimal point
\usepackage{bm}% bold math
\usepackage{gensymb}
\usepackage{feynmf}
\usepackage{lineno}
\usepackage{tikz-feynman}
\begin{document}

%\preprint{APS/123-QED}

%\title{Indication of a dynamically generated pentaquark configuration in the $\gamma n\rightarrow K^0_S\Sigma^0$ reaction}% Force line breaks with \\
%\title{Measurement of the $\gamma n\rightarrow K^0_S\Sigma^0$ differential cross section at around the $\K^*$ threshold}
%\title{Peak like structure observed in the $\gamma n\rightarrow K^0_S\Sigma^0$ differential cross section at BGOOD which appears consistent with a structure induced by a dynamically generated vector meson-baryon resonance}% Force line breaks with \\
\title{Measurement of the $\gamma n\rightarrow K^0\Sigma^0$ differential cross section over the $K^*$ threshold}% Force line breaks with \\
%\thanks{A footnote to the article title}%

\author{K. Kohl}\email{kohl@physik.uni-bonn.de}
%\corref{cor1}
\affiliation{Rheinische Friedrich-Willhelms-Universit\"at Bonn, Physikalisches Institut, Nu\ss allee 12, 53115 Bonn, Germany}

\author{T.C.~Jude}
\email{jude@physik.uni-bonn.de}
\affiliation{Rheinische Friedrich-Willhelms-Universit\"at Bonn, Physikalisches Institut, Nu\ss allee 12, 53115 Bonn, Germany}

\author{	S.~Alef}\thanks{No longer employed in academia}%%
\affiliation{Rheinische Friedrich-Willhelms-Universit\"at Bonn, Physikalisches Institut, Nu\ss allee 12, 53115 Bonn, Germany}
%\author{D. Bayadilov}
%\affiliation{ Rheinische Friedrich-Wilhelms-Universit\"at Bonn, Helmholtz-Institut f\"ur Strahlen- und Kernphysik, Nu\ss allee 14-16, 53115 Bonn, Germany}
%\affiliation{Petersburg Nuclear Physics Institute NRC "Kurchatov Institute", Gatchina, Leningrad District, 188300, Russia }
\author{ R.~Beck}%
\affiliation{ Rheinische Friedrich-Wilhelms-Universit\"at Bonn, Helmholtz-Institut f\"ur Strahlen- und Kernphysik, Nu\ss allee 14-16, 53115 Bonn, Germany}
%\author{A.~Bella}\thanks{No longer employed in academia}
%\affiliation{Rheinische Friedrich-Willhelms-Universit\"at Bonn, Physikalisches Institut, Nu\ss allee 12, 53115 Bonn, Germany}
%\author{J.~Bieling}\thanks{No longer employed in academia}
%\affiliation{Rheinische Friedrich-Willhelms-Universit\"at Bonn, Physikalisches Institut, Nu\ss allee 12, 53115 Bonn, Germany}
\author{ A.~Braghieri}%
\affiliation{ INFN sezione di Pavia, Via Agostino Bassi, 6 - 27100 Pavia, Italy}
\author{ P.L.~Cole}
\affiliation{Lamar University, Department of Physics, Beaumont, Texas, 77710, USA}
\author{D.~Elsner}
\affiliation{Rheinische Friedrich-Willhelms-Universit\"at Bonn, Physikalisches Institut, Nu\ss allee 12, 53115 Bonn, Germany}
\author{ R.~Di Salvo}%
\affiliation{INFN Roma ``Tor Vergata", Via della Ricerca Scientifica 1, 00133, Rome, Italy}
\author{ A.~Fantini}%
\affiliation{INFN Roma ``Tor Vergata", Via della Ricerca Scientifica 1, 00133, Rome, Italy}
\affiliation{Universit\`a di Roma ``Tor Vergata'', Dipartimento di Fisica, Via della Ricerca Scientifica 1, 00133, Rome, Italy}
\author{O.~Freyermuth}
\affiliation{Rheinische Friedrich-Willhelms-Universit\"at Bonn, Physikalisches Institut, Nu\ss allee 12, 53115 Bonn, Germany}
\author{ F.~Frommberger}%
\affiliation{Rheinische Friedrich-Willhelms-Universit\"at Bonn, Physikalisches Institut, Nu\ss allee 12, 53115 Bonn, Germany}
\author{ F.~Ghio}%
\affiliation{ INFN sezione di Roma La Sapienza, P.le Aldo Moro 2, 00185, Rome, Italy}
\affiliation{Istituto Superiore di Sanit\`a, Viale Regina Elena 299, 00161, Rome, Italy}
%\author{ S.~Goertz}%
%\affiliation{Rheinische Friedrich-Willhelms-Universit\"at Bonn, Physikalisches Institut, Nu\ss allee 12, 53115 Bonn, Germany}
\author{ A.~Gridnev}%
\affiliation{Petersburg Nuclear Physics Institute NRC "Kurchatov Institute", Gatchina, Leningrad District, 188300, Russia }
\author{D.~Hammann}\thanks{No longer employed in academia}
\affiliation{Rheinische Friedrich-Willhelms-Universit\"at Bonn, Physikalisches Institut, Nu\ss allee 12, 53115 Bonn, Germany}
\author{J.~Hannappel}
\affiliation{Rheinische Friedrich-Willhelms-Universit\"at Bonn, Physikalisches Institut, Nu\ss allee 12, 53115 Bonn, Germany}

\author{ N.~Kozlenko}
\affiliation{Petersburg Nuclear Physics Institute NRC "Kurchatov Institute", Gatchina, Leningrad District, 188300, Russia }
\author{ A.~Lapik}
\affiliation{ Russian Academy of Sciences Institute for Nuclear Research, Prospekt 60-letiya Oktyabrya 7a, 117312, Moscow, Russia}
\author{ P.~Levi Sandri}%
\affiliation{ INFN - Laboratori Nazionali di Frascati, Via E. Fermi 54, 00044, Frascati, Italy}
\author{ V.~Lisin}%
\affiliation{ Russian Academy of Sciences Institute for Nuclear Research, Prospekt 60-letiya Oktyabrya 7a, 117312, Moscow, Russia}
\author{ G.~Mandaglio}%
\affiliation{INFN sezione Catania, 95129, Catania, Italy}
\affiliation{Universit\`a degli Studi di Messina, Dipartimento MIFT,  Via F. S. D'Alcontres 31, 98166, Messina, Italy}
\author{ R.~Messi}%
\affiliation{INFN Roma ``Tor Vergata", Via della Ricerca Scientifica 1, 00133, Rome, Italy}
\affiliation{Universit\`a di Roma ``Tor Vergata'', Dipartimento di Fisica, Via della Ricerca Scientifica 1, 00133, Rome, Italy}
\author{ D.~Moricciani}%
\affiliation{ INFN - Laboratori Nazionali di Frascati, Via E. Fermi 54, 00044, Frascati, Italy}
\author{ V.~Nedorezov}\thanks{Deceased}%
\affiliation{ Russian Academy of Sciences Institute for Nuclear Research, Prospekt 60-letiya Oktyabrya 7a, 117312, Moscow, Russia}
\author{V.A~Nikonov}
\affiliation{ Rheinische Friedrich-Wilhelms-Universit\"at Bonn, Helmholtz-Institut f\"ur Strahlen- und Kernphysik, Nu\ss allee 14-16, 53115 Bonn, Germany}
\affiliation{Petersburg Nuclear Physics Institute NRC "Kurchatov Institute", Gatchina, Leningrad District, 188300, Russia }
\author{ D.~Novinskiy}%
\affiliation{Petersburg Nuclear Physics Institute NRC "Kurchatov Institute", Gatchina, Leningrad District, 188300, Russia }
\author{ P.~Pedroni}%
\affiliation{ INFN sezione di Pavia, Via Agostino Bassi, 6 - 27100 Pavia, Italy}
\author{ A.~Polonskiy}
\affiliation{ Russian Academy of Sciences Institute for Nuclear Research, Prospekt 60-letiya Oktyabrya 7a, 117312, Moscow, Russia}
\author{ B.-E.~Reitz}\thanks{No longer employed in academia}%
\affiliation{Rheinische Friedrich-Willhelms-Universit\"at Bonn, Physikalisches Institut, Nu\ss allee 12, 53115 Bonn, Germany}
\author{ M.~Romaniuk}
\affiliation{INFN Roma ``Tor Vergata", Via della Ricerca Scientifica 1, 00133, Rome, Italy}
\affiliation{Institute for Nuclear Research of NASU, 03028, Kyiv, Ukraine}
\author{ G.~Scheluchin}\thanks{No longer employed in academia}%%
\affiliation{Rheinische Friedrich-Willhelms-Universit\"at Bonn, Physikalisches Institut, Nu\ss allee 12, 53115 Bonn, Germany}
\author{ H.~Schmieden}%
\affiliation{Rheinische Friedrich-Willhelms-Universit\"at Bonn, Physikalisches Institut, Nu\ss allee 12, 53115 Bonn, Germany}
\author{ A.~Stuglev}%
\affiliation{Petersburg Nuclear Physics Institute NRC "Kurchatov Institute", Gatchina, Leningrad District, 188300, Russia }
\author{ V.~Sumachev}\thanks{Deceased}%
\affiliation{Petersburg Nuclear Physics Institute NRC "Kurchatov Institute", Gatchina, Leningrad District, 188300, Russia }
\author{ V.~Tarakanov}
\affiliation{Petersburg Nuclear Physics Institute NRC "Kurchatov Institute", Gatchina, Leningrad District, 188300, Russia }
%\author{V.~Vegna}
%\affiliation{Rheinische Friedrich-Willhelms-Universit\"at Bonn, Physikalisches Institut, Nu\ss allee 12, 53115 Bonn, Germany}%
%\author{T.~Zimmermann}
%\affiliation{Rheinische Friedrich-Willhelms-Universit\"at Bonn, Physikalisches Institut, Nu\ss allee 12, 53115 Bonn, Germany}
%\author{{\textcolor{red}{who do we include as authors?}}%_{}

% \email{Second.Author@institution.edu}
%\affiliation{%
% Authors' institution and/or address\\
% This line break forced with \textbackslash\textbackslash
%}%
%
%
%\author{{Charlie Author}
% \homepage{http://www.Second.institution.edu/~Charlie.Author}
%\affiliation{
% Second institution and/or address\\
% This line break forced% with \\
%}%
%\affiliation{
% Third institution, the second for Charlie Author
%}%
%\author{{Delta Author}
%\affiliation{%
% Authors' institution and/or address\\
% This line break forced with \textbackslash\textbackslash
%}%

%\cortext[cor1]{Corresponding author}
\collaboration{BGOOD Collaboration}%\noaffiliation

\date{\today}% It is always \today, today,
             %  but any date may be explicitly specified

\begin{abstract}{
	The differential cross section for the quasi-free photoproduction reaction $\gamma n\rightarrow K^0\Sigma^0$ was measured at BGOOD at ELSA from threshold to a centre-of-mass energy of \SI{2400}{MeV}. Close to threshold the results are consistent with existing data and  are in agreement with partial wave analysis solutions over the full measured energy range, with a large coupling to the $\Delta(1900)1/2^-$ evident.  This is the first dataset covering the $K^*$ threshold region, where there are model predictions of dynamically generated vector meson-baryon resonance contributions.  
	% An increase in the cross section is observed at forward angles above \SI{2000}{MeV}.
	%The data do not exclude the calculations of Ramos and Oset, which predicts a peak at around the $K^*$ threshold caused by interference with a resonance of dynamically generated vector meson-baryon states, however  no conclusions can be drawn without further data.
}

\end{abstract}

\pacs{Valid PACS appear here}% PACS, the Physics and Astronomy
                             % Classification Scheme.
%\keywords{Suggested keywords}%Use showkeys class option if keyword
                              %display desired
\maketitle

%\tableofcontents

\section{Introduction}\label{intro}

Associated strangeness photoproduction is a crucial tool to study nucleon resonance spectra.  A main motivation of the measurement of $KY$ channels over the last 15 years has been to search for \textit{missing resonances} which may only couple weakly to $N\pi$ final states\,\cite{Capstick:2000qj,Loring:2001kx}.  $K^0\Sigma^0$ has a threshold at 1690\,MeV, rendering the channel ideal to probe the third resonance region where many $s$-channel resonances up to high-spin states lie.  Studying this reaction is therefore a requirement to constrain phenomenological models and partial wave analyses (PWA) which attempt to describe the nucleon excitation spectrum of known resonances.  This includes PWA with dynamical coupled-channel frameworks\,\cite{Anisovich:2007bq,Anisovich:2014yza,CBELSATAPS:2019ylw,Ronchen:2018ury}, isobar models\,\cite{Skoupil:2016ast,Skoupil:2018vdh,Mart:1999ed, Clymton:2017nvp,Lee:1999kd,Janssen:2001wk,Janssen:2001pe,Janssen:2003zv}, and models incorporating Regge trajectories\,\cite{DeCruz:2011xi,DeCruz:2012bv,Bydzovsky:2019hgn} to fix $t$-channel contributions. 
% \textcolor{red}{Data close to threshold is of particular importance as the fewer contributing resonances reduces the number of parameters enabling an improved determination of hadronic couplings and form factors\,\cite{Mart:2014eoa}.}  
$K^0$ photoproduction data is also complementary to $K^\pm$ photoproduction as hadronic couplings can be related via isospin symmetry\,\cite{Mart:2011ez} and the absence of $t$-channel pseudo-scalar $K$ exchange ensures $s$-channel resonance contributions are more dominant (however there are still $K^*$ $t$-channel contributions).

%Multiquark mesonic and baryonic states beyond the 
%conventional  $q \bar q$ and $q q q $ valence quark configurations are a topical
%issue in contemporary hadron physics, and have been promoted by an abundance of $X, Y, Z$ mesons in the (hidden) charm sector\,\cite{Chen16,Liu19} and the discovery of the 
%$P_C$ baryon states\,\cite{Aaji15,Aaji19}, interpreted as pentaquark configurations \,\cite{Burns15}.  There are also well known issues in the light quark sector which defy conventional valence quark descriptions\,\cite{GR96}, for example, the mass-parity ordering of the N(1440)1/2$^+$ and the N(1535)1/2$^-$ resonances, and the $\bar{K}N$ molecular description of the $\Lambda(1405)$ supported by numerous measurements\,\cite{Zychor:2007gf, CLAS:2013rjt, BGOOD:2021sog}, Lattice QCD calculations\,\cite{Hall:2014uca,Molina:2015uqp} and chiral perturbation theory based models\,\cite{Oller:2000fj, Jido:2003cb, Oset:1997it, Hyodo:2011ur, Mai:2014xna}. 

Additionally, calculations based on vector meson-baryon interactions via coupled-channel unitary frameworks have predicted dynamically generated states contributing to $K^0\Sigma$ channels.   A model by Ramos and Oset\,\cite{RO13} explained a cusp-like structure observed in $K^0 \Sigma^+$ photoproduction\,\cite{Ewald12} at the $K^*$ threshold from the destructive interference between amplitudes containing $K^*\Lambda$ and $K^*\Sigma$ intermediate states, and magnified by a proposed $N^*(2030)$ vector meson-baryon dynamically generated resonance at the $K^* \Sigma$ threshold.  The model predicts that for photoproduction off the neutron, the interference of these amplitudes is constructive, resulting in a peak structure in the channel $\gamma n \rightarrow K^0\Sigma^0$.

%have been used to describe multiquark baryonic states in both the charm and light quark sectors, including  a prediction of the $P_C$ states prior to their discovery  by Wu \textit{et al.}\,\cite{Wu10}.  
%This resulted 

The complexity of identifying the $K^0\Sigma^0$ final state has led to a lack of data compared to the $K^+\Lambda$ and $K^+\Sigma$ channels, where the only available dataset is from the A2 Collaboration and covers the first 150\,MeV from threshold\,\cite{MAMI}.  Motivation for the study of  $K^0\Sigma^0$ photoproduction is therefore twofold.  Firstly, to constrain phenomenological models and PWA used to describe the nucleon excitation spectrum.  Secondly, to provide the first dataset over the $K^*$ threshold region in an attempt to discriminate between models including ``conventional" $s$-channel resonances and models predicting meson-baryon dynamically generated resonances beyond a  \textit{qqq} valence quark configuration.  This paper reports a measurement of the differential cross section of the reaction $\gamma n \rightarrow K^0\Sigma^0$ from threshold to 2400\,MeV, achieved with the BGOOD experiment\,\cite{technicalPaper} at the ELSA\,\cite{H06,H17} facility at Bonn University.

\section{Experimental setup and running conditions}\label{sec:setup}

%The BGOOD experiment \cite{technicalPaper} at the ELSA accelerator facility \cite{ELSA,ELSA2} in Bonn, Germany uses an energy tagged bremsstrahlung photon beam for meson photoproduction. 

The presented data was taken using an ELSA 
%For the presented data, an ELSA 
%the experiment was provided with an
electron beam of \SI{2.9}{GeV} incident upon a thin radiator
%\footnote{ Two different radiators were used, either diamond (\SI{560}{\mu m}) or copper (\SI{67}{\mu m}.)}
 to produce a collimated beam of bremsstrahlung photons.
The photon energies were determined by measuring the post bremsstrahlung electron momenta in the \textit{Photon Tagger}.
Two data taking periods with an \SI{11}{cm} long target containing either liquid deuterium or hydrogen with identical beam conditions were used.  The hardware trigger, which was the same for both data taking periods, required a signal in the Photon Tagger and an energy deposition of at least \SI{200}{MeV} in the BGO Rugby Ball. Details on the characterization and modelling of the trigger are in Ref.\,\cite{Alef:2020yul}. The signal was extracted from the deuterium data and the hydrogen data was used to subtract background originating off the the proton in the deuterium.
%were used, trigger and beam conditions were the same for both.
%Two data taking periods using either a \SI{11}{cm} long liquid deuterium or hydrogen target were used.
%  which was converted to a real photon beam via bremsstrahlung.
% The \SI{11}{cm} long target contained either liquid deuterium or hydrogen. 
The integrated photon flux from threshold to a centre-of-mass energy\footnote{To correctly determine the centre-of-mass energy the Fermi-momentum of the target particle needs to be known, however since this was not possible event by event, the target was assumed to be at rest.} of \SI{2400}{MeV} was $6.39\cdot 10^{12}$ and $5.78\cdot10^{12}$, respectively.

%%It is ideally suited to measure strangeness photo-production.
% The electron beam is converted to a real photon beam via bremsstrahlung. By measuring the electron momentum in the Photon Tagger, the photon energy is determined.
%%  The target is surrounded by a plastic scintillator barrel and two Multi-Wire-Proportional Chambers vetoing particle charge and for track reconstruction.
 The BGOOD experiment is composed of a magnetic forward spectrometer complemented with a central BGO calorimeter\,\cite{technicalPaper}.
  The \textit{BGO Rugby Ball}, is composed of 480 BGO crystals, which surround the target and cover a polar angle range of $25\degree<\theta_{\mathrm{lab}}<155\degree$. Each crystal spans $6\degree$ to $10\degree$ in the polar angle $\theta_{\mathrm{lab}}$, $11.25\degree$ in the azimuthal angle $\phi$ and has a depth of $\SI{24}{cm}$, corresponding to 21 radiation lengths. The time resolution of \SI{2}{ns} allows for a clean separation of multiple photons for neutral meson reconstruction. A minimal required energy deposition of \SI{1.5}{MeV} per individual crystal and \SI{25}{MeV} per crystal cluster suppresses neutron background.
  A barrel type arrangement of plastic scintillators interior to the BGO Rugby Ball is utilized to veto charged particles.
  The forward spectrometer covers the angular range $1\degree<\theta_{\mathrm{lab}}<12\degree$. A series of tracking detectors
   sandwiching the  open dipole magnet\,\cite{technicalPaper} are used 
%  At forward directions ($1\degree<\theta_{\mathrm{lab}}<12\degree$) a series of tracking detectors and an open dipole magnet are used
for charged particle identification and momentum reconstruction.
%   and time of flight walls
%   two tracking detectors (MoMo, SciFi) in front of an open dipole magnet and eight driftchambers behind are used for momentum reconstruction, and together with three time-of-flight walls, particle identification. 
The small intermediate range is covered by a segmented array of 96 plastic scintillators for charged particle detection.
%The small intermediate range is covered by segments of 90 plastic scintillators, which are also used for charged particle detection.

%A detailed description of the experiment can be found at Ref.\,\cite{technicalPaper}. 

\begin{figure}
	% 	\centering
	\includegraphics[width=.49\textwidth]{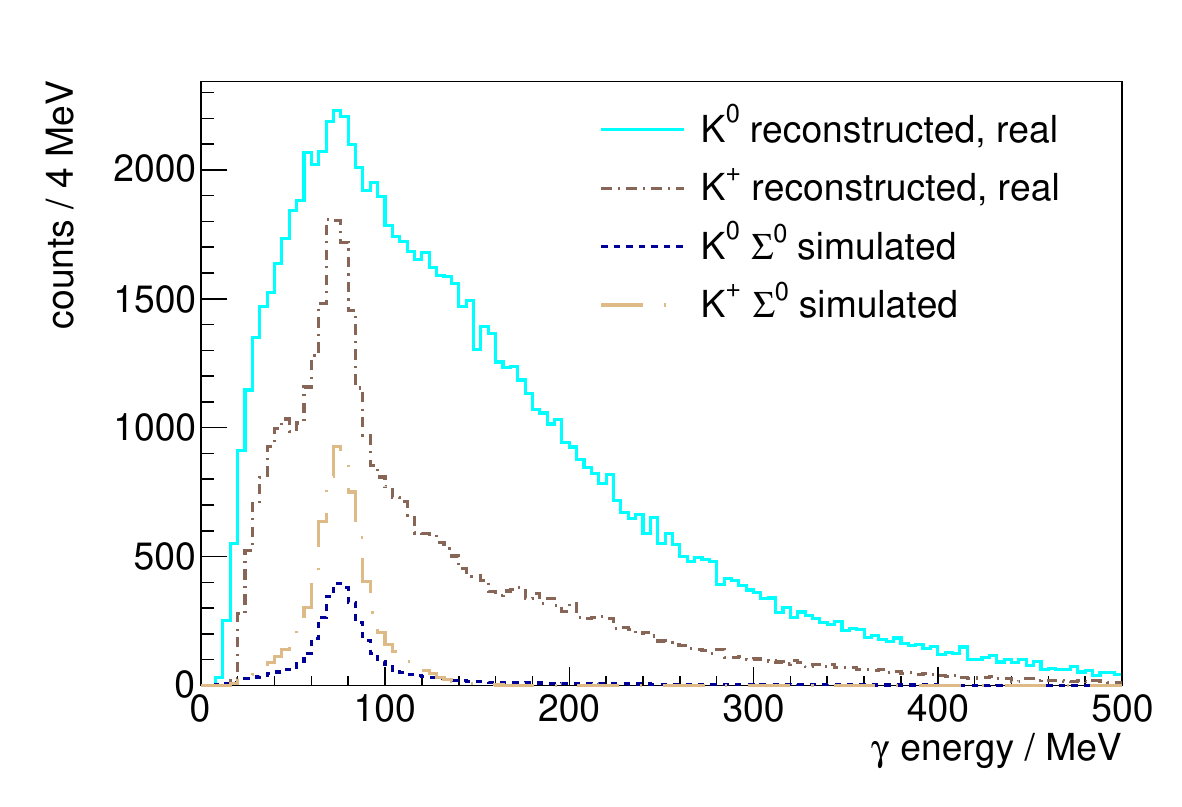}
	\caption{Photon energy spectrum in the \mysigma{} rest frame for real and simulated data for the channels $\gamma p\rightarrow K^+\Sigma^0$ and $\gamma n\rightarrow K^0\Sigma^0$ at an arbitrary scale. 
		% 		. As a prove of the method, the channel $K^+$\mysigma{} is shown.
	}
	\label{fig:decay_photon}       % Give a unique label
\end{figure}

\section{Selection of $\gamma n \rightarrow K^0\Sigma^0$ events}\label{sec:selectevents}
\kaon{} candidates were identified via the decay ${K^0_S\rightarrow \pi^0\pi^0\rightarrow (\gamma\gamma)(\gamma\gamma)}$ in the BGO Rugby Ball. 
Two photon pairs were required where the invariant masses were within
 $\SI{30}{MeV/c^2}$ of the \pion{} mass, which corresponds to $\pm2\sigma$.
Three additional selection criteria were used to isolate the reaction channel. First, the missing mass to the \kaon{} candidates was required to be consistent with the \mysigma{} mass, lying between 1150 and \SI{1250}{MeV/c^2}. 
%Since the Fermi momentum is not known for every event, the missing mass is calculated assuming a neutron at rest. This results in a slight broadening of energy and momentum resolution.
%Secondly, the photon from the decay $\Sigma^0\rightarrow \gamma\Lambda$ was identified.
Secondly, identification of the photon from the decay $\Sigma^0\rightarrow \gamma\Lambda$ was required.
%Thirdly 
% events where the topology is consistent with $\Sigma^0\rightarrow \gamma\Lambda\rightarrow\gamma\pi^-p$.
%In the \mysigma{} rest frame this photon from the two body \mysigma{} decay is fixed at \SI{75}{MeV}.
%has an energy of approximately the mass difference between \mysigma{} and $\Lambda$ of $\approx$\SI{75}{MeV}. This is used to select signal events and suppress background.
To achieve this, photons which were not identified as a \pion{} decay photon were boosted into the rest frame of the \mysigma.
%To identify this photon a fifth \mygamma{} that was not included as a \pion{} decay photon was boosted into the rest frame of the \mysigma.
Fig.\,\ref{fig:decay_photon} shows the energy of the boosted \mygamma{} in this frame, where a peak at \SI{75}{MeV} from the two-body \mysigma{} decay is expected.
 A small peak is visible over a large background at this energy, which is consistent with the simulated data. The decay photon from the channel $\gamma p \rightarrow K^+$\mysigma{} is also shown, where the signal is cleaner\footnote{This technique was used to identify the $K^+\Sigma^0$ channel in Ref.\,\cite{decayphoton}.}.
% and the peak can be separated from the background much better.
%  All three curves show the peak at the same expected position. 
Events were selected where the photon energy was between $\SI{54}{MeV}$ and \SI{96}{MeV}, corresponding to $\pm3\sigma$. 
%  A cut was then applied at $\pm\SI{21}{MeV}$ around \SI{75}{MeV}. 
%  The effect of this can be seen in Fig.\,\ref{fig:overview_invmass}, a bump is now visible at the \kaon{} mass over a large amount of background.

  \begin{figure}
 	%	\centering
 	\includegraphics[width=.49\textwidth]{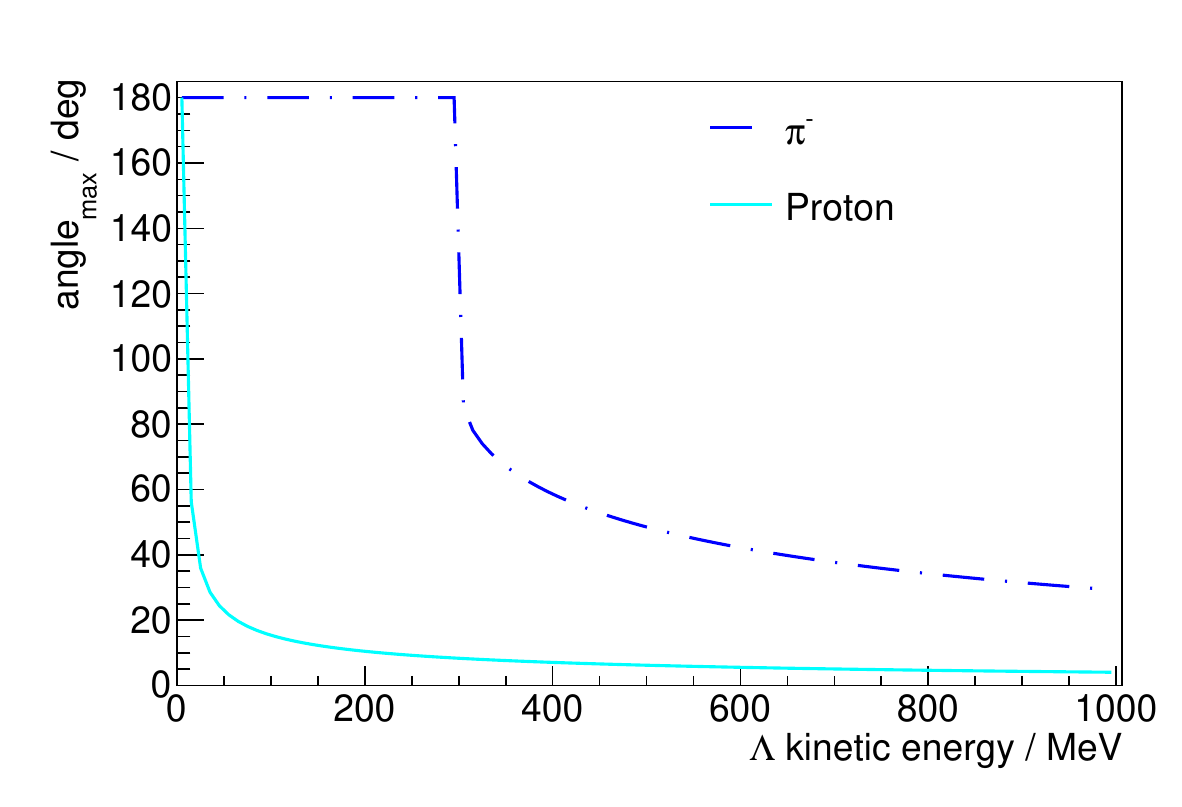}
 	\caption{Maximum possible angle between either the proton or the $\pi^-$ and $\Lambda$ as a function of $\Lambda$ kinetic energy in the decay $\Lambda\rightarrow p\pi^-$.}
 	\label{fig:lambda_angles}       % Give a unique label
 \end{figure}

%\subsection{$\Lambda$ decay particles}
The final selection criterion required the detection of exactly
%To further reduce the background events were selected where
 two charged particles consistent with the decay $\Lambda\rightarrow\pi^-p$.
 The $\Lambda$ momentum was calculated from the missing momentum to the \kaon{} and the photon from the \mysigma{} decay, and the angle between the charged particles and the $\Lambda$ was determined. This angle must lie within the kinematically allowed region shown in Fig.\,\ref{fig:lambda_angles}. 
Nearly all $\pi^-$ and $p$ were detected in the BGO Rugby Ball or SciRi, where no completely clean identification between them could be made.  All events where at least one charged particle was within the angular limit for the proton and the other for the $\pi^-$ were therefore retained.  The small angle allowed between the $\Lambda$ and proton in particular, provided a vital constraint for removing other misidentified reaction channels.  An additional 10\degree{} was permitted to account for detector resolution and the unmeasured Fermi-momentum of the target nucleon within the deuteron.
%Finally the missing mass to \kaon{} was selected to be around the \mysigma{} mass between 1150 and \SI{1250}{MeV/c^2}.

\begin{figure}
	%	\subcapraggedrighttrue
	%	\centering
	\includegraphics[width=0.49\textwidth]{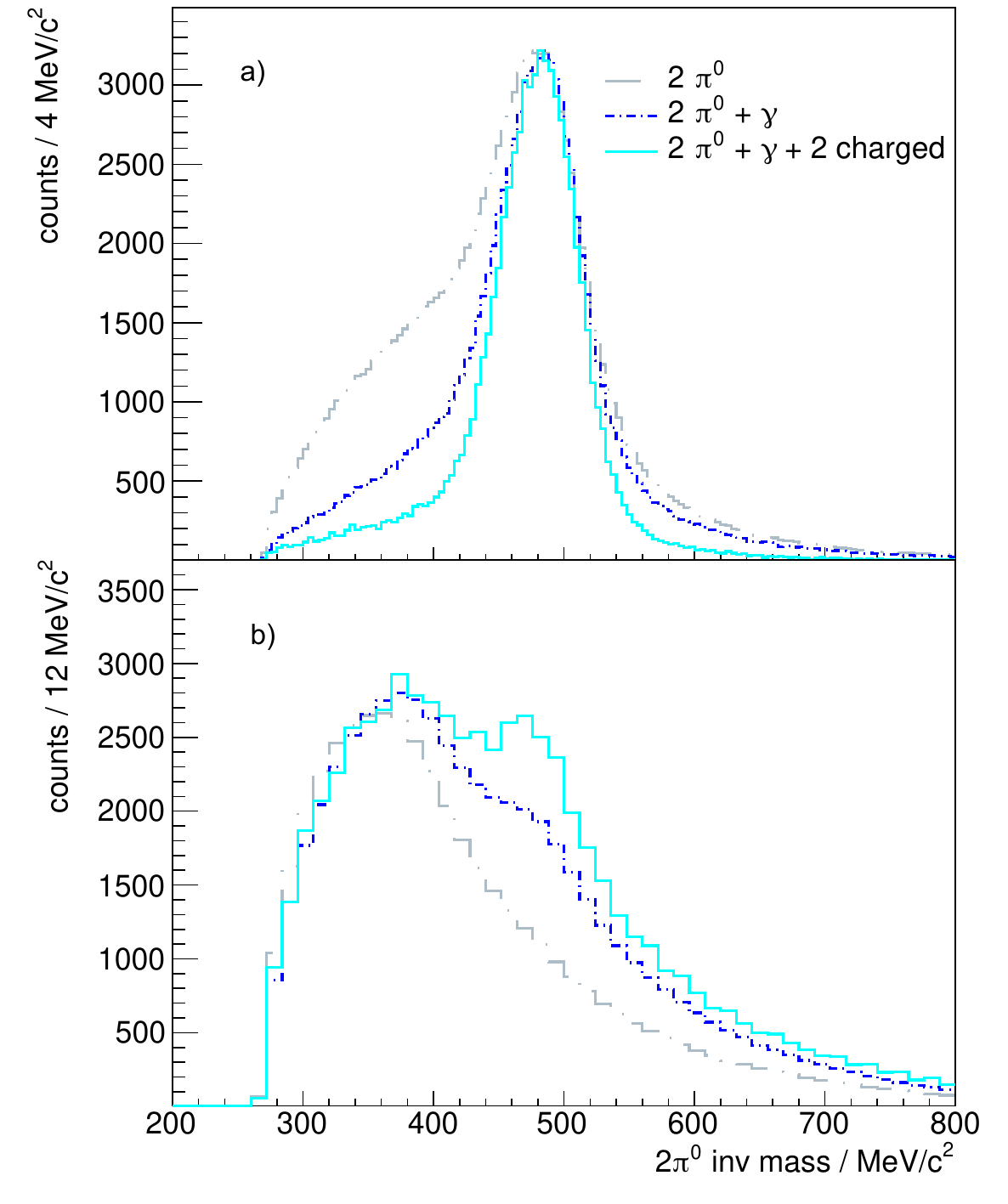}
	
	%}
	%		The first two are scaled by a factor 0.0045 and 0.0035 respectively.
	\caption{The invariant mass distribution of the 2\pion{} system for all measured $W$ and $\cos(\theta)$ intervals after different selection criteria for (a) simulated $\gamma n \rightarrow K^0\Sigma^0$ and (b) real data using a deuterium target.  The selection criteria are: only two \pion{}, two \pion{} and \mygamma{} from \mysigma{} decay, two \pion{},\mygamma{} and two charged particles from $\Lambda$ decay. The grey and dark blue distributions are scaled to approximately match the maximum height.}
	\label{fig:overview_invmass}       % Give a unique label
\end{figure}

For a given event, all combinations of particle assignment to the five neutral and two charged particles were retained if they passed the selection criteria.  No kinematic fit was applied as the lack of energy information of the detected charged particles combined with the unknown target Fermi-momentum prevented a full four-momentum constraint.  Fig.\,\ref{fig:overview_invmass} shows how the different selection cuts affect the invariant mass distribution of the two \pion{} system.
 Fig.\,\ref{fig:overview_invmass}(a) depicts simulated $\gamma n\rightarrow K^0\Sigma^0$ data, where increasing the selection criteria removes the low mass shoulder originating from combinatorial background, for example misidentifying a $\gamma$ from the \mysigma{} decay as a \pion{} decay \mygamma.  This background is small after including all selection criteria and was estimated as contributing a systematic uncertainty of 3\,\%.
Fig.\,\ref{fig:overview_invmass}(b) depicts real data using a deuterium target, where the peak corresponding to the \kaon{} invariant mass becomes increasingly pronounced from the background distribution with increased selection criteria.

%	the spectrum for real data, Fig.\,\ref{fig:twopiinvmasssim} shows the same spectrum for MC simulated $\gamma n\rightarrow K^0\Sigma^0$ events. The latter illustrates the effect of the different selections on combinatorial background. After all selections this amounts to approximately 3\%.}
% shows the effect of the different cuts on the invariant mass of two \pion{} using the deuterium target dataset.
%the $\Lambda$ decay particles are identified. Here the charged decay to $\pi^-p$ is used. There is no complete particle identification, any two charged particles are accepted. However kinematics give some constrains on the possible angles. Fig.\,\ref{fig:lambda_angles} shows the maximum possible angle of proton and $\pi^-$ in the $\Lambda$ rest frame. To account for resolution effects an extra 10\degree{} are allowed. 
%As shown in Fig.\,\ref{fig:overview_invmass}, this drastically reduces statistics, but also improves the signal to background ratio a lot.

\begin{figure}
%	\centering
	\includegraphics[width=.49\textwidth]{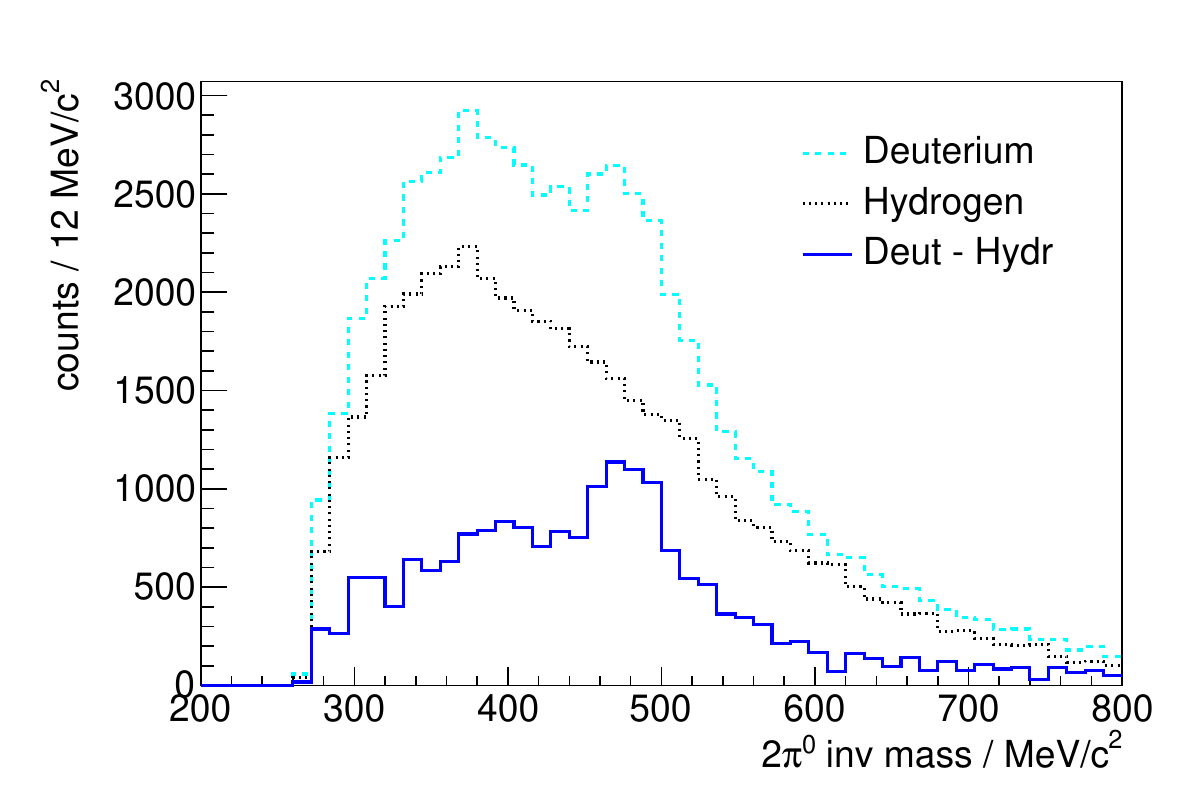}
	\caption{The invariant mass distribution of the  2\pion{} system after all selection criteria for deuterium and hydrogen data. The hydrogen data is scaled by luminosity and subtracted from the deuterium data.}
	\label{fig:hydrogen}       % Give a unique label
\end{figure}

%\subsection{Proton contamination}
%Since the liquid deuterium target consists of protons and neutrons, and the reaction investigated here is only happening on the neutron, every reaction happening on the protons of the target is background. This knowledge helps removing this background.
Background from reaction channels off the proton in the deuteron was subtracted by applying the same selection criteria to the liquid hydrogen target dataset. The hydrogen data was scaled according to the ratio of the photon beam flux and target densities.
% The analysis for the deuterium is performed assuming a fixed target neutron. 
 To account for the broadening of momentum and mass distributions caused by the unmeasured Fermi motion, the proton target four-momentum
 in the computation of kinematic quantities was smeared according to the momentum distribution of nucleons in deuterium\,\cite{Lacombe80}.
% To get the correct amount of hydrogen data, it is scaled with the ratio of the different flux and densities and finally subtracted from the deuterium data. 
 Fig.\,\ref{fig:hydrogen} shows the scaled hydrogen data together with the deuterium data and the resulting spectrum after subtracting hydrogen data from deuterium data. 
% Differences in the shape outside the signal region result from background channels specific to the neutron, that are not present in the hydrogen data.

%\subsection{Fitting to data}
Two methods were used to fit to the $K^0\Sigma^0$ signal and remaining background.  The first used simulated phase-space distributions of the dominant background channels, referred to later as \textit{PS}.  Simulated data  was used to determine the fraction of background channel events passing the selection criteria.  The dominant channels were found to be multi-pion production ($\gamma n \rightarrow 3\pi N$ and $\gamma n \rightarrow 4\pi N$) and $\gamma n \rightarrow  \eta n$, all of which gave almost identical $2\pi^0$ invariant mass spectra.  The required topology of five neutral and two charged particles was satisfied by these background channels either exactly, or by missing a particle in small acceptance gaps in the experimental setup, or by falsely identifying an additional particle due to split-off clusters in the BGO Rugby Ball and particles scattering off detector components.  The channel $\gamma n \rightarrow 3\pi^0n$ was chosen as representative of the multi-pion channels and used to describe the background distribution. Other channels were found to provide negligible contributions.

The second method, later referred to as \textit{RD}, used real data  to describe the background. To generate this distribution, a \kaon{} candidate and an additional photon were required, however this photon was not identified as a \mysigma{} decay photon candidate and there was no selection criteria on charged particle multiplicity or topology.
% found and the first of the addidational selection criteria were fullfilled, but there was no \mysigma{} decay photon candidate, herein referred as \textit{RD}.
\begin{figure}
	%	\centering
	%	\fontsize{1}{12}\selectfont
	%\resizebox{0.49\textwidth}{!}{\input{./graphics/Fits.tex}}
	\includegraphics[width=.48\textwidth]{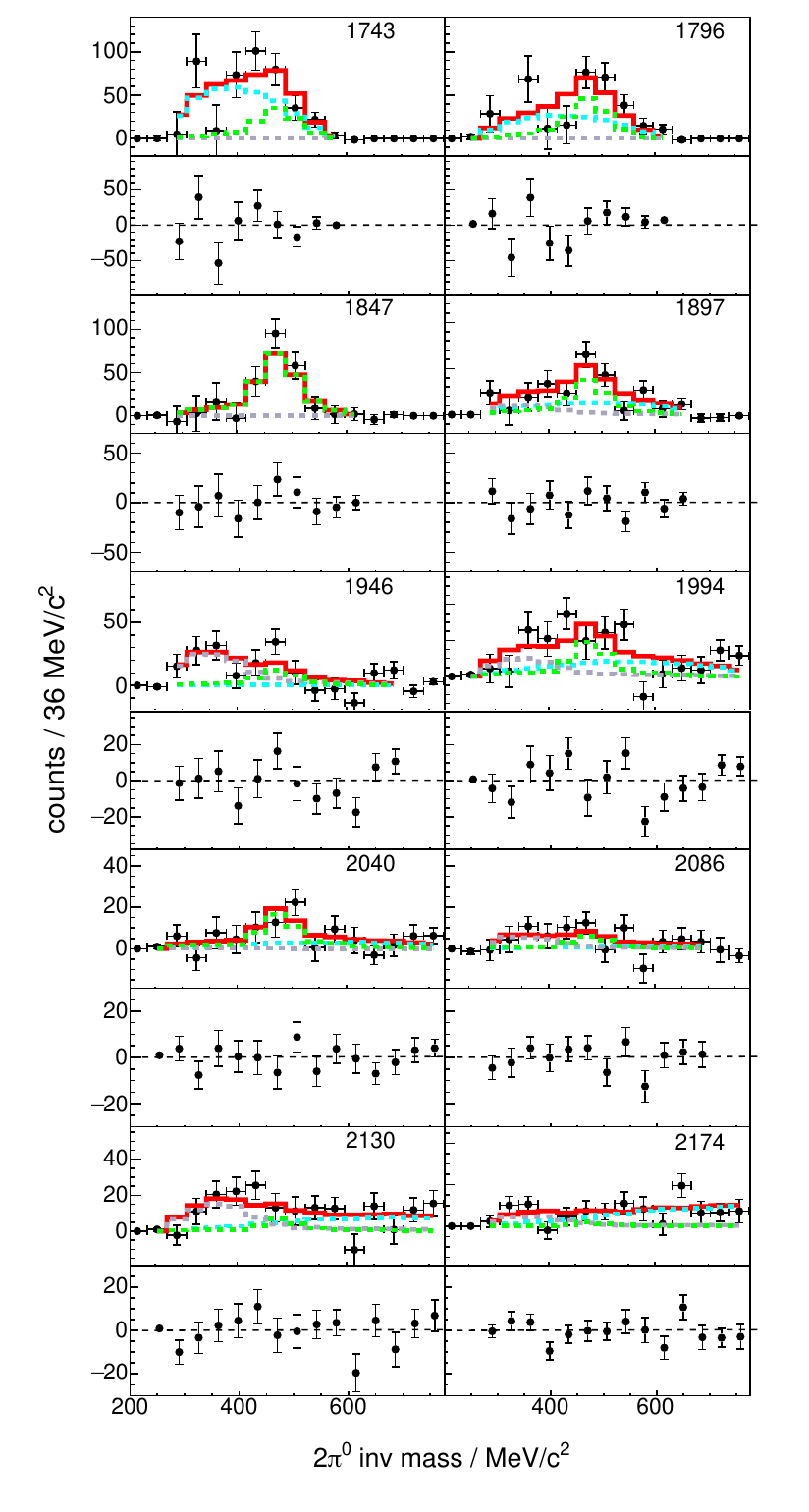}
	\caption{Example of fits to the 2\pion{} invariant mass spectrum. The angular region is show from $0.2<$\kaonangle$<0.5$ in 10 energy bins. The centre of each energy bin is given as $W$ in MeV in the top right corner. Black points are measured data, the signal is shown in green. The two background channels ${\gamma n \rightarrow 3\pi^0n}$ and $\gamma n \rightarrow  \eta n$ are shown in light blue and gray. The sum of all channels is depicted as a solid red line. The residuals are plotted under each fit.
		%		 the dashed line indicates 0. 
		%		The center of each energy bin is given as $E_\gamma$ in MeV in the top right corner. 
		%		Shown are both fitting methods for two different energy bins (RD left, PS right). 
		%		\textcolor{red}{this is just one fitting method and the most forward bin, but I think it's enough to show the fits are working. Have to adjust y-axis for plots in one line} 
	}
	\label{fig:fit}       % Give a unique label
\end{figure}

 \begin{figure}
	%	\centering
	\includegraphics[width=.46\textwidth]{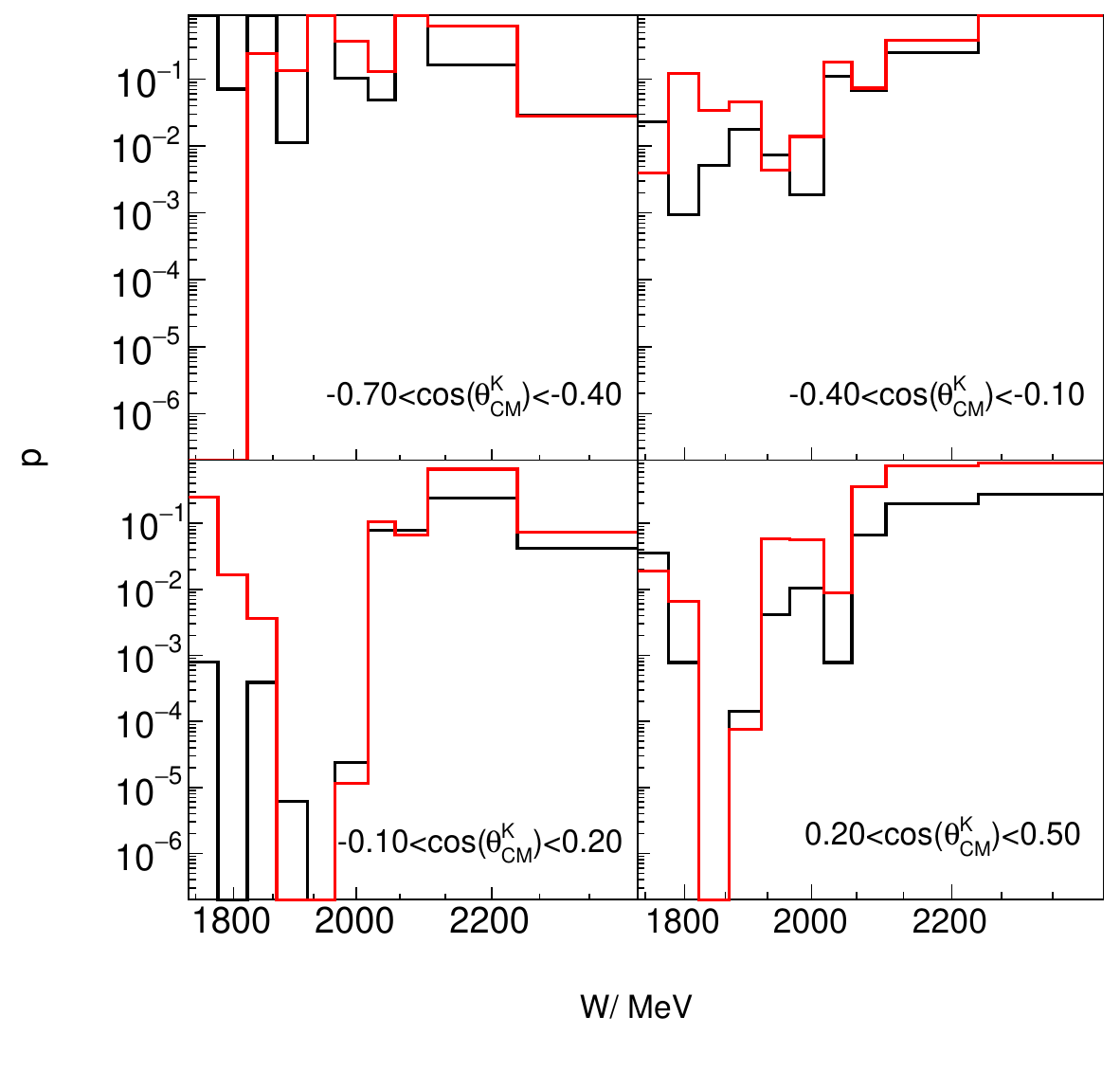}
	\caption{$p$ as a function of $W$ for the two different background distributions. The black line corresponds to a background distribution using PS description, the red line corresponds to a background distribution using RD description respectively.
		%		\textcolor{red}{If first energy bin is excluded, same y-axis can be chosen for all plots and this will look a lot nicer}
	}
	\label{fig:hypothesis}       % Give a unique label
\end{figure}

In both cases, the signal shape was phase-space generated using simulated \kaon\mysigma{} data and a full Geant4\,\cite{Geant4} simulation of the experimental setup. Roofit\,\cite{Verkerke:2003} was used to fit the data with a maximum likelihood fit. Fig.\,\ref{fig:fit} shows fit examples using the PS background description in the angular range $0.2<$\kaonangle$<0.5$, where \kaonangle{} is the cosine of the centre-of-mass polar angle of the \kaon.

The limited statistical precision reduces the usefulness of $\chi^2$ distributions. Instead a hypothesis test was performed to prove the necessity of the simulated signal distribution to describe the data. 
The test gives the probability of the data following a given distribution which is only comprised of background, the hypothesis of which is denoted $H_0$.
%The test gives the probability of the data following a given distribution under the hypothesis $H_0$ this distribution being only background.
 This is achieved by creating 10000 Monte Carlo (MC) samples from the background distributions each with the same statistical precision as the real data distributions.
 Each MC sample is fitted twice. The first fit is with background (BG) only, the second is with background and signal (BG+S).
 $\zeta^2$, given in Eq.\,\ref{zeta2} is calculated for each fit, where $N_{\text{fit}}$ and $ N_{\text{data}}$ are the number of events in each bin of the fitted spectrum for the fitted function and the data respectively, and $\Delta N_{\text{data}}$ is the corresponding error.
 \begin{align}
 	\zeta^2=\sum_{\text{bins}}\left( \frac{N_{\text{fit}}-N_{\text{data}}}{\Delta N_{\text{data}}}\right)^2
 	\label{zeta2}
 \end{align}
%  For each fit $\zeta^2$ is calculated:

  For each MC sample the difference is calculated:
  \begin{align}
  \Delta \zeta^2 = \zeta^2(BG)-\zeta^2(BG+S)
  \end{align}
  This is repeated for the real data, denoted $ \Delta \zeta^2_{\text{real}}$. The distribution of $\Delta \zeta^2$ for the 10000 MC samples under the hypothesis $H_0$ is denoted $g(\Delta \zeta^2| H_0) $.
  
%  and fitting each once with background only(BG) and once with background and signal(BG+S). For each sample $\Delta \zeta^2 = \zeta^2(BG)-\zeta^2(BG+S)$ is calculated where $\zeta^2=\sum_{bins}\left( \frac{N_{fit}-N_{data}}{\Delta N_{data}}\right)^2$. The distribution of $\Delta \zeta^2$ under the hypothesis $H_0$ is called $g(\Delta \zeta^2| H_0) $. 
  A measure of agreement with $H_0$ can then be calculated from Eq.\ref{eq:pvalue}:
  \begin{align}
  p= \int_{\Delta \zeta^2_{\text{real}}}^{\infty}g(\Delta \zeta^2| H_0)
  \label{eq:pvalue}
  \end{align}
%  $p= \int_{\Delta \zeta^2_{real}}^{\infty}g(\Delta \zeta^2| H_0) $.
   Fig.\,\ref{fig:hypothesis} shows $p$ using the two background descriptions PS and RD. Both descriptions 
%   gives this probability of the measured data resulting from the PS and RDS background description only, where they
    generally agree with each other. 
    $p$ is low where the signal gives a significant contribution to the fitted spectrum,
%     a signal could be fitted,
      indicating that a background distributions alone is not sufficient to describe the data.
      
      An alternative method to separate signal and background was made using side band subtraction techniques\,\cite{Kohl:2022vfr}. The resulting yields were in agreement to the fitting methods described above.
      
%A Kolmogorov test \cite{Kolmogorov33} was used to prove the necessity of the simulated signal distribution to describe the data. This test gives the probability of two distributions being identical. In Fig.\,\ref{fig:hypothesis} the ratio of these probabilities for fits with and without the signal included is shown.
%The fact that this ratio is clearly above one in the signal region indicates that the signal is real.
%one or above for all kinematic regions, demonstrates the importance of including the signal to describe the data.
% A value above one indicates the fit including the signal describes the data better than the fit without.
%Fig.\,\ref{fig:fit_NS} shows an example of a fit with and without the simulated signal distribution, where it is clear the background alone is not sufficient.
%  can not describe the data.

Fitting to the \kaon{} invariant mass does not discriminate between $\gamma n \rightarrow K^0\Sigma^0$ and $\gamma n \rightarrow K^0\Lambda$, however, the selection criteria strongly suppressed the contribution from $\gamma n \rightarrow K^0\Lambda$.
%Simulated data was used to determine the small fraction of $K^0_S\Lambda$ events passing these selection criteria.
%A small fraction of background $K^0_S\Lambda$ events remained in the event yield, simulated data was used to determine the percentage of events passing the selection criteria.
% This was typically 0.05\% and 0.015\% at beam energies \SI{1250}{MeV} and \SI{1750}{MeV} respectively.
Simulated data and cross section measurements from Ref.\,\cite{lambda} were subsequently used to determine and subtract the remaining contribution of $K^0\Lambda$ to the $K^0\Sigma^0$ yield. This was a small contribution and largest near threshold, with for example 16\,\% and 2\,\% of the extracted yield in the angular bin $0.2<$\kaonangle$<0.5$ at center-of-mass energies \SI{1796}{MeV} and \SI{2040}{MeV} respectively.

\begin{figure}
	%	\centering
	\includegraphics[width=.49\textwidth]{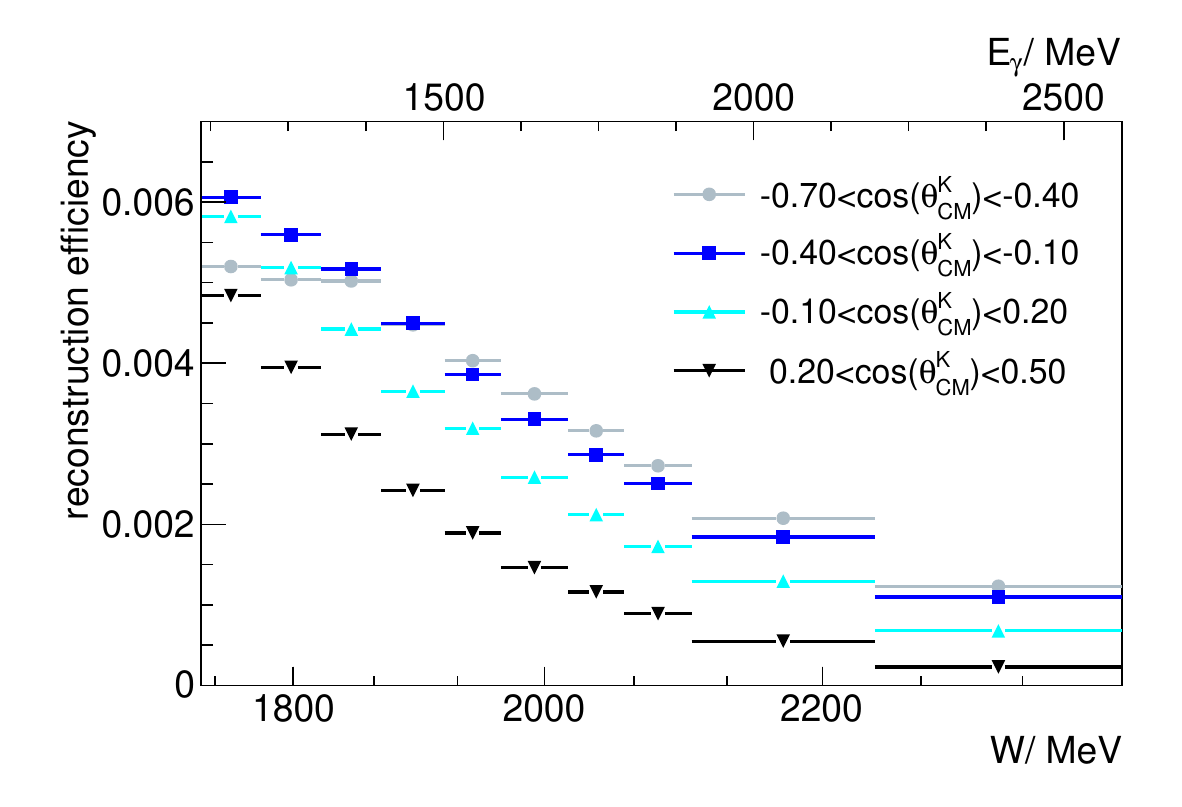}
	\caption{Reconstruction efficiency as a function of energy for four different \kaonangle{} intervals.}
	\label{fig:recoeff}       % Give a unique label
\end{figure}

\begin{table}
	\begin{tabular}{l c}
		\hline
		\hline
		Source & \% error\\
		\hline
		Photon flux& 4\\
		Target length & 1\\
		Beam energy calibration & 1\\
		Modelling of hardware triggers & 1\\
		$\pi^0$ identification & 3\\
		$\Sigma^0 \rightarrow \gamma \Lambda$ identification  & 6\\
		Selection of the missing ($\Sigma^0$) mass&  3\\
		Charged particle identification & 4\\
		Combinatorial background & 3\\
		Subtraction of hydrogen background & 5\\
		\kaon$\Lambda$ subtraction & 1\\
		%		Fitting to signal and background & 3 \\
		\hline
		Summed in quadrature & 11\\
		\hline
		\hline
	\end{tabular}
	\caption{Sources and values of systematic uncertainties (not including the fitting systematic uncertainty).}
	\label{tab:systematics}
	
\end{table}

The reconstruction efficiency is depicted in Fig.\,\ref{fig:recoeff} as a function of energy for four different \kaonangle{} ranges. This includes the branching ratios of the $K^0$ eigenstates $K^0_S$ and $K^0_L$ and the detected $K^0_S$ and $\Lambda$ decay modes, which limits the efficiency to $\approx$10\,\%. Requesting 5 neutral particles in the central calorimeter further reduces the efficiency to below $1\,\%$. No structures are seen that could cause artefacts in the measured cross section.

\begin{figure*}[]
	%	\centering
	\includegraphics[width=1.05\textwidth]{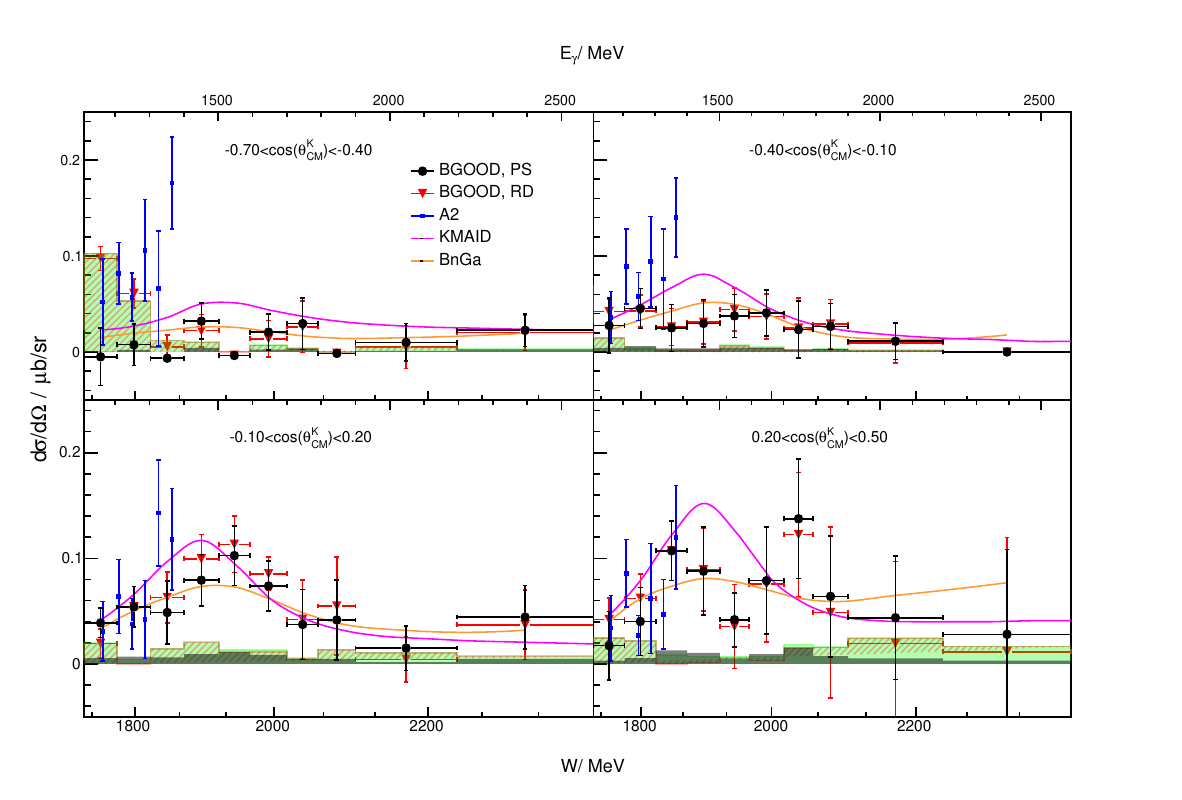}
	\caption{$\gamma n \rightarrow K^0\Sigma^0$ differential cross section as a function of $W$ in four bins in \kaonangle{} for both fitting methods RD and PS(red triangles and black circles respectively).  Vertical error bars are the statistical error, horizontal error bars indicate the bin width. Systematic errors are split into \textit{scaling errors} shown as grey columns and \textit{fitting errors} shown as dashed red columns. The sum in quadrature of both is shown as green columns. Data from Akondi \textit{et al.} (A2 Collaboration)\,\cite{MAMI} is shown in blue squares, with median values of \kaonangle{} of -0.625, -0.375, 0.125 and 0.375. Calculations from Kaon MAID\,\cite{Lee:1999kd} and BnGa\,\cite{BnGa} are shown as magenta and orange lines respectively. 
		%The predicted total cross section from Ramos and Oset\,\cite{RO13} is included in the interval $0.2<$\kaonangle$<0.5$, denoted by the red axis on the right, but at an arbitrary scale relative to the measured differential cross section.
	}
	\label{fig:cs}       % Give a unique label
\end{figure*}

 Table \ref{tab:systematics} shows the systematic uncertainties. The identification of the photon from the decay $\Sigma^0\rightarrow \gamma\Lambda$ and the subtraction of hydrogen background
 %The photon flux normalization, \pion{} identification and the selection of the missing (\mysigma) mass 
 are the dominating uncertainties at 6\,\% and 5\,\% of the measured cross section, respectively. The uncertainty on the photon flux normalization was determined as explained in Ref.\,\cite{Alef:2020yul}. Systematic uncertainties specific to this analysis were estimated by varying the selection criteria at each step and determining the effect on the extracted cross section. The systematic uncertainty of fitting is determined as the difference between the cross section of the two methods to fit the background (RD and PS). While all other systematic uncertainties are a constant fraction of the measured cross section and therefore can only change the global scaling of the dataset, the fitting uncertainties permit point to point fluctuations of the data points. These uncertainties are therefore shown separately in Fig.\,\ref{fig:cs}.
 
\section{Results}\label{sec:results}
The differential cross section for ${\gamma n \rightarrow K^0\Sigma^0}$ is shown in Fig.\,\ref{fig:cs} as a function of energy in four bins in \kaonangle.
The two methods used to describe and subtract background show a good agreement, with the exception of the most backward angle bin, $-0.7<$\kaonangle$<-0.4$, where there is a discrepancy of up to \SI{0.1}{\mu b/sr} in the first two energy bins from threshold to $W=$\SI{1823}{MeV}. This is due to limited phase space in the $2\pi^0$ invariant mass spectrum preventing a clean separation of signal and background.
The data of Akondi \textit{et al.} (A2 Collaboration)\,\cite{MAMI} are shown as the blue squares from threshold to $W=$ \SI{1855}{MeV}.  When combining the statistical and systematic uncertainties of both datasets (the systematic uncertainty of the A2 data varies from 0.001 to 0.004\,$\mu$b/sr), there is reasonable consistency over most of the kinematic coverage, however the A2 data generally appears higher.  This is most pronounced at the two most backward angles at $W  = $\SI{1855}{MeV}, where there is a discrepancy of approximately $1\sigma-2\sigma$ beyond the combined uncertainties.

Calculations from the Kaon MAID effective Lagrangian model\,\cite{Lee:1999kd} and the Bonn-Gatchina Partial Wave analysis (BnGa)\,\cite{BnGa} are shown as the magenta and orange lines respectively.  The BnGa calculation includes dominant contributions from $S_{11}$ and $P_{11}$ partial waves and gives an agreement to the data over the full measured \kaonangle{} range. The Kaon MAID calculation also appears to have a reasonable agreement in the two most forward \kaonangle{} intervals, whereas in the two backward intervals the calculation lies approximately between this data and the A2 data.  The Kaon MAID model includes resonant contributions from  $\Delta(1650)1/2^-$, $N(1710)1/2^+$, $N(1720)3/2^+$,  $\Delta(1900)1/2^-$ and $\Delta(1910)1/2^+$.  The peak at $W = 1900$\,MeV observed in the data most prominently in the interval $-0.10 <$\kaonangle{} $< 0.20$ is described by the large coupling to the $\Delta(1900)1/2^-$. 

%\textcolor{green}{A discrimination between the different models will only be possible with improved statistical precision.}
  
%In the most forward \kaonangle{} interval, a structure may become visible above \SI{2000}{MeV}, which within limited statistical precision is consistent with the model of Ramos and Oset\,\cite{RO13} of the vector-meson baryon dynamically generated state which described the cusp observed in $K^0\Sigma^+$ photoproduction.  Fig.\,\ref{fig:cs} shows a superposition of the model result in the $0.2<$\kaonangle$<0.5$ interval. This is the total cross section integrated over the full \kaonangle{} range and arbitrarily scaled for a comparison with the measured data.\footnote{A differential calculation is currently not available.}  However, the current statistical precision does not permit a reliable conclusion.

 The model by Ramos and Oset\,\cite{RO13} predicted a peak at the $K^*$ threshold caused by a vector-meson baryon dynamically generated state.
This dataset does not exclude a structure at ${W\approx2040}$\,MeV and $0.20 <$\kaonangle{} $< 0.50$, however the current statistical precision does not permit a  conclusion and further data is required to discriminate between phenomenological models in this energy range.	
	
%It is interesting to note the model calculation by Ramos and Oset\,\cite{RO13} including a vector-meson baryon dynamically generated state at the $K^*$ threshold.

%A comparison with the model of Ramos and Oset\,\cite{RO13} is not done here, as a differential calculation is currently not available. However, in the most forward \kaonangle{} interval, a structure is not excluded, which, within limited statistical precision, might be consistent with the model of Ramos and Oset\,\cite{RO13} of the vector-meson baryon dynamically generated state which described the cusp observed in $K^0\Sigma^+$ photoproduction.}

%The distribution of the Ramos and Oset model differs significantly from the Kaon MAID and BnGa calculations, which will enable 
%A discrimination between the different models will be possible, when improved statistical precision is available. 
Contributions from final state interactions can not be disregarded without additional studies, however calculations for quasi-free photoproduction off the deuteron of $K^+Y$\,\cite{KplusFSI, Salam:2004gz} show them to be negligible over the kinematic range presented here.

\section{Conclusions}\label{sec:summary}

A first measurement of the reaction $\gamma n \rightarrow K^0 \Sigma^0$ is presented from threshold to 2400\,MeV, spanning the region of the $K^*$ threshold. 
The channel was identified via the decays ${K^0_S\rightarrow\pi^0\pi^0}$ and $\Sigma^0\rightarrow\Lambda\gamma\rightarrow(p\pi^-)\gamma$ at the BGOOD experiment. Two different methods were used to describe background from other reaction channels passing the selection criteria, which after fitting to $2\pi^0$ invariant mass spectra to extract the signal proved compatible.  The data are consistent with existing data and well described by PWA solutions.
%, however a structure above 2000 MeV is not excluded, which is consistent with a model describing the cusp in $K^0\Sigma^+$ photoproduction via dynamically generated meson-baryon states. 
Improved statistical precision and an extension to more forward \kaonangle{} is required to fully characterize and discriminate between model calculations.

In BnGa no resonances are put in a priori. Instead poles emerge from the  K Matrix formalism within specific partial waves. What is included here is S11 and P11 partial waves.
%\vspace{0.9cm}
 \section*{Acknowledgements}

We thank the staff and shift students of the ELSA accelerator for providing an excellent beam. Fruitful discussions with E. Oset and A. Ramos are gratefully acknowledged. We thank S. Neubert for his help with hypothesis testing. We also thank A. Sarantsev for providing calculations from the Bonn-Gatchina Partial Wave analysis.

 H. Schmieden thanks for the support from the Munich Institute for Astro- and Particle Physics (MIAPP), funded by the DFG under Germany's Excellence Strategy - EXC-2094-390783311. P.L. Cole gratefully acknowledges the support from both the U.S. National Science Foundation (NSF-PHY-1307340, NSF-PHY-1615146, and NSF-PHY-2012826) and the Fulbright U.S. Scholar Program (2014/2015). This work was supported by the Deutsche Forschungsgemeinschaft Project Numbers 388979758 and 405882627, the RSF grant number 19-
42-04132, the Third Scientific Committee of the INFN and
the European Union's Horizon 2020 research and innovation programme under grant agreement number 824093. 

%\textcolor{red}{is there more support, from russian/italian side?}
\bibliography{References}
\bibliographystyle{science}

\end{document}